\documentclass[utf8]{frontiersFPHY}
\usepackage{url,microtype}
\usepackage[onehalfspacing]{setspace}
\usepackage[utf8]{inputenc}
\usepackage{amsmath}
\usepackage{amsfonts}
\usepackage{amssymb}
\usepackage{graphicx}

\def\keyFont{\fontsize{8}{11}\helveticabold }
\def\firstAuthorLast{David J Miller} 
\def\Authors{John McDowall, David J Miller$^{*}$}
\def\Address{SUPA, School of Physics and Astronomy, University of Glasgow, \\
						Glasgow, G12 8QQ, United Kingdom}
\def\corrAuthor{David J Miller}
\def\corrEmail{David.J.Miller@Glasgow.ac.uk}

\begin{document}
\onecolumn
\firstpage{1}

\title[The Multiple Point Principle]{The Multiple Point Principle and Extended Higgs Sectors}

\author{\Authors}
\address{}
\correspondance{}
\extraAuth{}

\maketitle

\begin{abstract}
The Higgs boson quartic self-coupling in the Standard Model appears to become zero just below the Planck scale, with interesting implications to the stability fo the Higgs vacuum at high energies. We review the {\em Multiple Point Principle} that suggests the quartic self-coupling should vanish exactly at the Planck scale. Although this vanishing is not consistent with the Standard Model, we investigate Higgs sectors extended with additional states to test whether one may satisfy the high scale boundary condition while maintaining the observed Higgs mass. We also test these scenarios to ensure the stability of the vacuum at all energies below the the Planck scale and confront them with experimental results from the LHC and Dark Matter experiments.

\tiny
 \keyFont{ \section{Keywords:} Higgs boson, Beyond the Standard Model, Extended Higgs sectors, Renormalization Group, Multiple Point Principle, Dark Matter} %
\end{abstract}

\section{Introduction}
\label{sec:introduction}
It is widely believed that investigations of the Higgs boson and the resulting breaking of Electroweak Symmetry provide the best opportunity for finding new physics beyond the Standard Model (SM). In part, this is because the Higgs boson is the most recently discovered fundamental particle \cite{Aad:2012tfa}, and investigations of its properties are still underway (though so far no significant deviation from the SM has been observed \citep{Aad:2015zhl, Khachatryan:2014kca, Khachatryan:2014jba, Aad:2015gba}). This view is reinforced by the required relative smallness of the Higgs boson mass and its related {\em hierarchy problem}. Since the SM Higgs boson mass is unprotected by any symmetries, it should have large quantum corrections of magnitude comparable to the scale of new physics. To restore a physical Higgs mass of order the Electroweak scale one must {\em fine-tune} to ensure the unnatural cancelation of the bare Higgs mass with its corrections. Provided there is new physics of some type beyond the SM (a reasonable assumption, given its large number of problems and omissions) this is a genuine and very real issue that must be addressed.

The combined ATLAS and CMS value of the Higgs mass \citep{Aad:2015zhl}, $m_h = 125.09 \pm 0.23\,$GeV, raises further questions. This is a challenging value for both supersymmetry and composite Higgs models, requiring a significant tuning of parameters or a non-minimal field content \citep{Buttazzo:2013uya,Craig:2013cxa,Ross:2017kjc}, making it difficult to motivate any particular models and unclear which direction to head next. However, this particular mass has another reason for being peculiar - it is just the right value to allow the Higgs potential to be {\em metastable} at high energies \cite{Buttazzo:2013uya}.

\begin{figure}[htb]
  \centering
  \includegraphics[width=\textwidth]{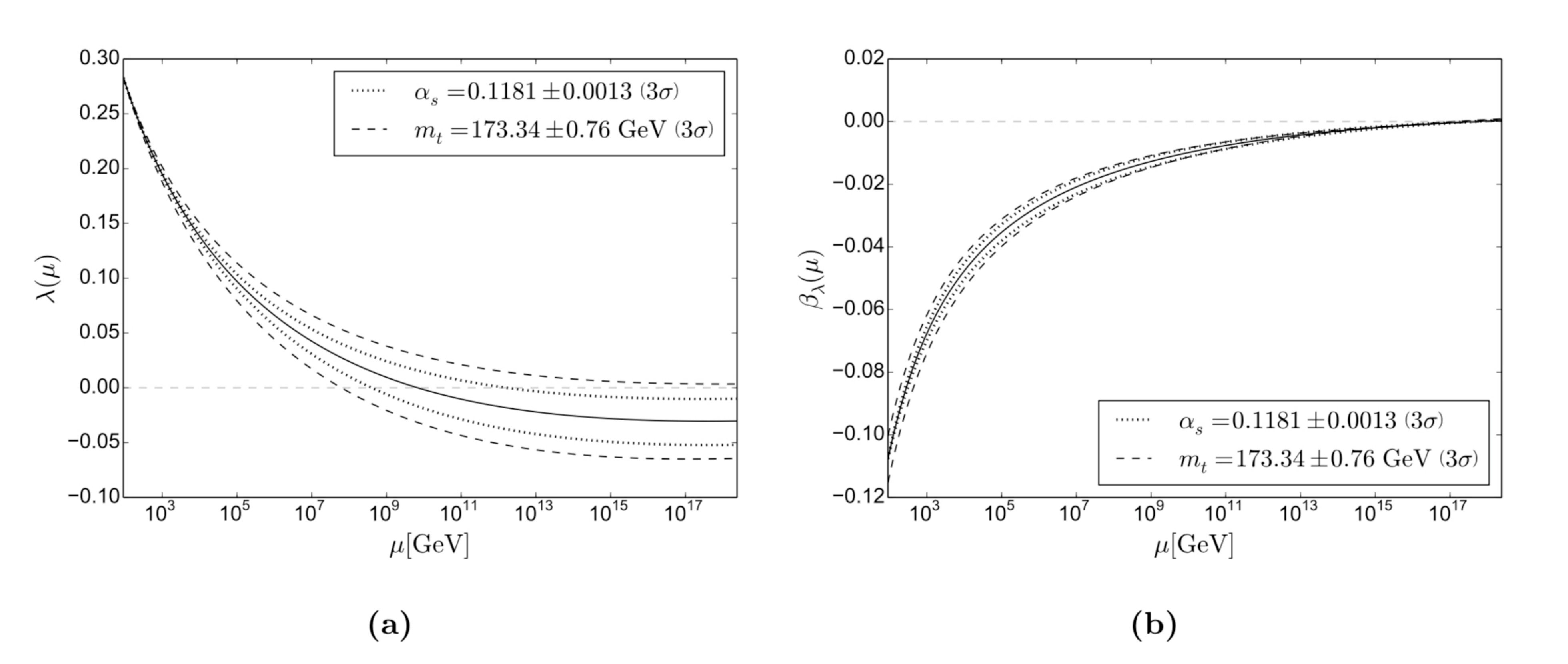}
  \caption{Three-loop running of the SM Higgs quartic coupling $\lambda$ and its $\beta$ function with $3 \, \sigma$ uncertainties from the top pole mass $m_t$ (dashed) and the strong coupling constant $\alpha_s$ (dotted). These plots originally appeared in Ref.~\cite{McDowall:2018tdg}. \label{fig:SM_lambda_beta}}
\end{figure}

As usual for a parameter of a Quantum Field Theory, the Higgs quartic coupling $\lambda$ evolves with energy according to the Renormalization Group (RG) and  is pulled downwards at higher energies by the large top-quark mass. If it were to run to negative values the potential may become unstable and the correct pattern of Electroweak Symmetry breaking is lost. Indeed, requiring absolute stability of the vacuum up to the Planck scale $M_{\rm Pl}$, i.e.\ $\lambda(M_{\rm Pl}) \ge 0$, places a limit on the top mass \cite{Buttazzo:2013uya},
\begin{equation}
\label{eq:SM_topmassstabilitylimit}
m_t < 171.36 \pm 0.46 \text{ GeV},
\end{equation}
which is in tension with the current experimental value by about $2.6 \, \sigma$. Figure~\ref{fig:SM_lambda_beta}(a) shows the quartic coupling dependence on renormalization scale $\mu$, and the $3 \, \sigma$ uncertainties that arise from the uncertainties in the top-quark mass $m_t$ and the strong coupling constant $\alpha_s$. The quartic coupling turns negative at an energy scale of $\mu \sim 10^{10}$ GeV, though a stable potential is not ruled out due to the uncertainties. However, a very small negative value is not a catastrophe, since the vacuum may still be {\em metastable} with a lifetime much longer than the age of the universe. That nature should choose this metastable vacuum is intriguing. Why does the quartic coupling become so very nearly zero right at the Planck scale?

We may gain further insight by examining the beta-function of the quartic coupling, shown in  Figure~\ref{fig:SM_lambda_beta}(b). As indicated already in Figure~\ref{fig:SM_lambda_beta}(a), the running of $\lambda$ flattens out at high energies, i.e.\ $\beta_{\lambda}(M_{\rm Pl}) \approx 0$ too. We stress that in the SM this is {\em not} an ultraviolet fixed-point since $\lambda$ would continue to evolve if we increased the energy further. However, if some new physics theory takes over above the Planck scale, then the SM running becomes irrelevant and we must instead consult the new theory. If this new theory sets $\lambda = \beta_\lambda=0$ at the Planck scale we may recover a low energy phenomenology very similar to what we observe, modulo the slight deviation in the Higgs mass.

In this article, we will review one proposed high scale possibility, the {\em Mutliple Point Principle} (MPP) \cite{Froggatt:1995rt}. Although this is not compatible with the SM running, it provides a Higgs mass prediction that is curiously close to the measured value. We will then examine several theories with extended Higgs sectors to see if they alter the running sufficiently to provide the correct Higgs mass. For recent investigations of alternative high scale boundary conditions at $M_{\rm Pl}$ see, for example, Refs.~\citep{Degrassi:2012ry,Holthausen:2011aa,Iacobellis:2016eof,Eichhorn:2014qka,Khan:2014kba,Helmboldt:2016mpi}.

\section{The Multiple Point Principle in the SM}

The Multiple Point Principle (MPP) asserts that nature chooses the Higgs potential parameters so that different phases of electroweak symmetry breaking may coexist. This is analagous to how ice, water and vapour may coexist for specific values of temperature and pressure near water's triple-point. Since the two phases must be energetically comparable in order to coexist, this means that the potential should have at least two degenerate vacua, that is an additional vacuum degenerate with the usual Electroweak vacuum.

The authors of this principle argue in Ref.~\cite{Froggatt:1995rt} that this is rather natural if we consider {\em extensive} variables constrained by some new physics theory at high energies, as long as the system has a rather strong first order phase transition. Again we may use the analogy of water and note that slush (in which ice and liquid water coexist) is present for a (relatively) wide range of extensive variables (in this case temperature and pressure) due to the existence of a first order phase transition. Returning to the Higgs potential, a possible {\em extensive} quantity could be $\langle |\phi|^2 \rangle$. If this were set by some new physics theory at the Planck scale with a strong first order phase transition, it would be rather likely to find $\langle |\phi|^2 \rangle\sim M_{\rm Pl}^2$, leading to a second degenerate vacuum at the Planck Scale. In essense, this principle is relying on a rather flat distribution of {\em extensive} parameter space set at the Planck scale matching to a rather peaked distribution of {\em intensive} parameters (i.e.\ the usual Higgs potential parameters) due to a strong first-order phase transition, which in turn leads to a second degenerate vacuum \cite{bennett:1996hx}

We should note that what this Planck scale theory could be is still unknown, and Ref.~\cite{Froggatt:1995rt} makes no attempt to describe one, using only general principles to support the assertion. Also, we note that this provides no explanation of why the Planck scale is so much bigger than the electroweak scale. Nevertheless, the contraints on the Higgs parameters does provide a {\em prediction} of the Higgs boson mass that can be compared with experiment, and we further note that this prediction was first made long before the Higgs boson discovery.

\begin{figure}[htb]
  \centering
  \includegraphics[width=0.5
\textwidth]{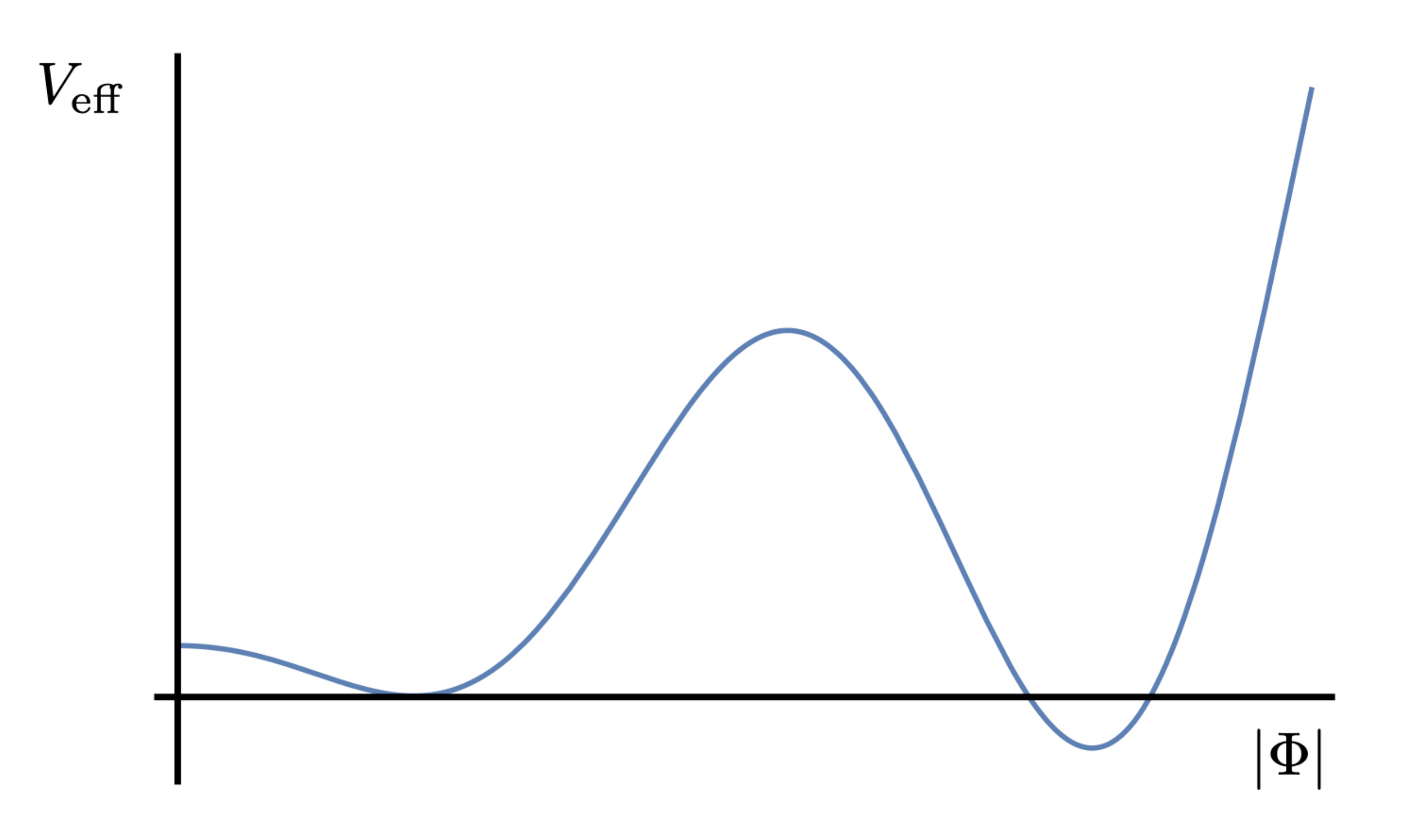}
  \caption{A schematic depiction of the one-loop effective potential in the SM. This is intended only to present a general picture of the minima and is not to scale. \label{fig:SMpot}}
\end{figure}

The one-loop Coleman-Weinberg effective potential~\cite{Coleman:1973jx} can be written,
\begin{equation}
	V_{\rm eff} = - \mu^2(\mu) \phi^2 + \frac{1}{4} \lambda \left( \mu \right) \phi^4(\mu) + \frac{1}{16\pi^2} V_1,
\end{equation}
where $V_1$ takes the schematic form $V_1 \sim \phi^2 \log \left( \phi^2/\mu^2 \right)$. For a more explicit form see, for example, Ref.~\cite{Jegerlehner:2014xxa}. We see that at one-loop, in addition to the new logarithmic contribution, the parameters $\mu$ and $\lambda$ become energy dependent. For low field values (and low energies) this reproduces the usual ``wine-bottle'' potential of the Higgs mechanism, but for higher field values, the logarithm pulls the potential back down. Eventually the $\phi^4$ terms becomes dominant and the potential will remain stable at the Planck scale if $\lambda(M_{\rm Pl})>0$. However, the additional structure causes a second minimum very close to the Planck scale. This is schematically depicted in Figure~\ref{fig:SMpot}. In the SM, taking the measured central values of the Higgs potential parameters, the second vacuum is of slightly lower energy than the Electroweak vacuum, causing the potential to be metastable. The MPP posits that the two minima should be degenrate.

For high field values the effective potential is dominated by its quadratic term, $V_{\rm eff} \approx \lambda \left( \mu \right) \phi^4$, so the second minima at the Planck scale requires
\begin{equation}
\label{eq:SM_MPP}
\frac{d V_{\rm eff}}{d \phi} \bigg\vert_{\phi=M_{\rm Pl}} \approx  \lambda \left( \mu \right) \phi^3 + \frac{1}{4} \beta_{\lambda} \left( \mu \right) \phi^4 = 0.
\end{equation}
We see that the MPP is satisfied if $\lambda(M_{\rm Pl}) = \beta_{\lambda}(M_{\rm Pl}) =0$.

Applying this boundary condition, the MPP hypothesis gave an early prediction~\cite{Froggatt:1995rt} of the Higgs mass $m_h = 135 \pm 9$ GeV, which is remarkably good considering it was made 17 years before the discovery of the Higgs boson, and they simultaneously predicted the top-quark mass (finding $173 \pm 5\,$GeV) in the same year it was discovered. A more recent calculation using the measured top-quark mass and newer determinations of e.g.\ $\alpha_s$, gave $m_h = 129 \pm 1.5$ GeV \cite{Buttazzo:2013uya}. Although this is slightly too high to be compatible with our by now very accurate Higgs mass measurement, it is still rather remarkable.

\begin{figure}[!tbp]
\centering
\includegraphics[width=\textwidth]{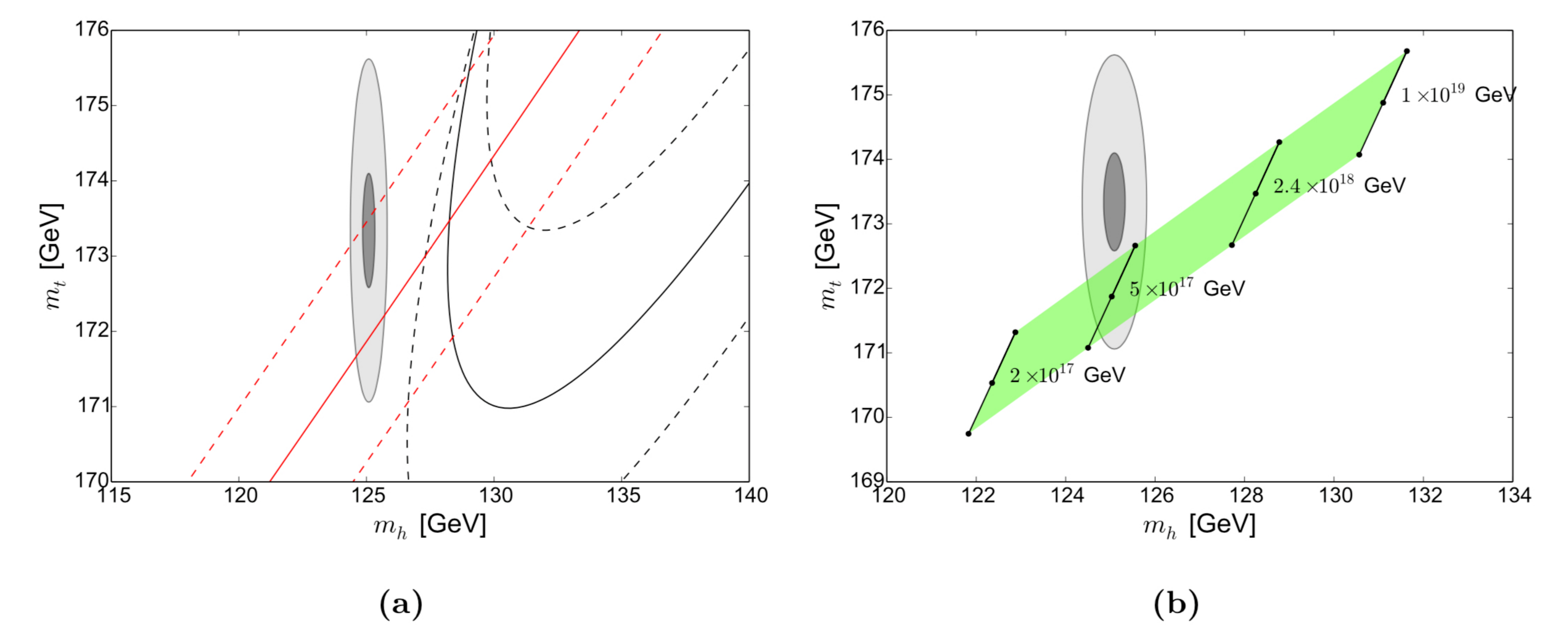}
\caption{\textbf{(a)} $\lambda \left( M_{Pl} \right) = 0$ (red) and $\beta_{\lambda} \left( M_{Pl} \right) = 0$ (black) contours in the $m_h-m_t$ plane. The dashed lines show $3 \,\sigma$ variations in $\alpha_s \left( M_Z \right) = 0.1181 \pm 0.0013$. \textbf{(b)} Mass values that satisfy both boundary conditions at various UV scales. The green region corresponds to a $1\, \sigma$ uncertainty in $\alpha_s$. Ellipses show the experimentally allowed values of $m_t$ and $m_h$ with $1 \,\sigma$ (dark grey) and $3\,\sigma$ (light grey) uncertainty. These plots originally appeared in Ref.~\cite{McDowall:2018tdg}.
\label{fig:SM_lambdabetalambda}}
\end{figure}

Figure \ref{fig:SM_lambdabetalambda}(a) shows contours corresponding to the boundary conditions $\lambda(M_{\rm Pl})=0$ and $\beta_{\lambda} \left( M_{Pl} \right) = 0$ in the $m_h-m_t$ plane, and we see that a slighly heavier Higgs is needed for both conditions to be satisfied. These contours are calculated using three-loop SM RG equations; the Higgs mass is calculated to two-loop order, while the top mass additionally contains three-loop QCD corrections. This plot is in agreement with the similar plot in Ref.~\citep{Degrassi:2012ry}, but we used a different value of the uncertainty in the strong coupling constant $\alpha_s \left( M_Z \right) = 0.1181 \pm 0.0013$ to reflect more recent estimates \cite{PDG:2015}. We also use the reduced Planck scale $M_{\rm Pl} = 2.4 \times 10^{18}\,$GeV as our scale at which these boundary conditions are set. Figure \ref{fig:SM_lambdabetalambda}(a) shows that $\lambda \left( M_{\rm Pl} \right) = 0$ can be satisfied with an acceptable value of $m_h$ for a top mass $171\, {\rm GeV} \lesssim m_t \lesssim 174\,$GeV, and although the corresponding value of $\beta_{\lambda} \left( M_{\rm Pl} \right)$ is not zero, it is extremely small.

Note that we have required that these boundary conditions be satisfied at $M_{\rm Pl}$, but if the theory that dictates the appearance of a second minimum were to become active at a lower energy scale, these boundary conditions would need to be altered. Figure \ref{fig:SM_lambdabetalambda}(b) shows the $m_h-m_t$ plane with points that satisfy both boundary conditions $\lambda = \beta_{\lambda} = 0$ simultaneously at different UV scales. The green region corresponds to a $1\, \sigma$ uncertainty in $\alpha_s$. We see it is possible to obtain a Higgs mass that is within experimental limits by applying these boundary conditions at approximately $5 \times 10^{17}$ GeV. It's interesting to note that this is a scale of importance in string scenarios (see e.g \citep{Ginsparg:1987ee, Witten:1996mz}).

As one approaches the Planck Scale, one might expect gravity to become significant and contribute to the RGE running of couplings. The study of these effects has caused some confusion in the literature. An initial calculation of the effect on the running of gauge couplings \cite{Robinson:2005fj}, using a quantised Einstein-Hilbert action as an effective field theory below the Planck scale, showed that this alters the gauge couplings sufficiently to render them asymptotically free. However, this calculation was disputed \cite{Pietrykowski:2006xy,Toms:2007sk} on the grounds that the derived result is gauge-dependent and therefore unreliable; a calculation performed with a different gauge choice (the harmonic gauge) instead revealed the contributions to be exactly zero. A recalculation was then done using the gauge-invariant background field method \cite{Toms:2010vy,Mackay:2009cf} and found a result in support of the original claim that the gauge coupling is rendered asymptotically free, though with a modified $\beta$-function. Also see Refs.~\cite{He:2010mt,Daum:2009dn} for alternative calculations. Calculations have also been performed to asses the affect on the quartic Higgs self-coupling relevant to the MPP \cite{Haba:2014qca,Wang:2015oka}. These two calculations disagree on the sign of the gravitational contributions to Yukawa couplings, but the corrections to the predicted Higgs mass are small; they predict a Higgs mass of ``approximately 130 GeV'' and ``$\gtrsim 131.5$ GeV'', neither of which are differing very far from the earlier prediction of $129 \pm 1.5$ GeV \cite{Buttazzo:2013uya} and remain incompatible with the SM. See also Ref.~\cite{Branchina:2018xdh} for a discussion of the effect on the electroweak vacuum of Planck suppressed operators.


\section{A Real Singlet Extension}

The simplest extension to the Higgs sector is to include an extra real singlet $S$, with potential,
\begin{eqnarray}
V \left( \Phi, S \right) &=& \mu^2 \Phi^{\dagger} \Phi + m_S^2 S^2 + \lambda \left( \Phi^{\dagger} \Phi \right)^2 + \lambda_S S^4
 + k_2 \Phi^{\dagger} \Phi S^2,
\label{eq:real_potential}
\end{eqnarray}
where a $Z_2$ symmetry, under which the new scalar is odd, has been used to eliminate terms odd in $S$ (see Ref.~\cite{Robens:2015gla} for a discussion of this model). During electroweak symmetry breaking, the real singlet field can acquire a non-zero vacuum expectation value (vev) $v_S$ alongside the SM Higgs. The usual Higgs scalar may then mix with the new singlet, though this mixing should not be too strong if we want to avoid LHC constraints. The singlet mass $m_S$ is fixed by the tadpole equation minimising the potential, analagous to the fixing of $\mu$ using the vev $v$. This leaves the parameters $\lambda$, $\lambda_S$, $k_2$ and $v_S$. We refer to this as the ``broken phase''. Alternatively, if the new scalar does not aquire a vev (i.e.\ $v_S=0$) the tadpole equation becomes trivial and cannot be used to remove $m_S$. Therefore we have parameters $\lambda$, $\lambda_S$, $k_2$ and $m_S$. Now the scalars do not mix, and the new scalar may be a Dark Matter candidate, so we refer to this as the ``Dark Matter phase''.

This real singlet model has been investigated in the context of the MPP in Refs.~\citep{Haba:2013lga,Haba:2014sia,Haba:2016gqx,Hamada:2014xka,Kawana:2014zxa,Kawana:2015aca}, with varying results. Haba et al.\ \cite{Haba:2013lga} investigated the model in the Dark Matter phase for the MPP as well as the Veltman condition \cite{Veltman:1980mj}. They found that both boundary conditions could be accommodated (separately) with a 126 GeV Higgs boson, while simultaneously providing the correct DM relic density. An alternative approach was taken in Refs.~\cite{Haba:2014sia,Haba:2016gqx} where the MPP was instead imposed on the real singlet model with the addition of an extra right-handed neutrino. Again, the MPP could be made compatible with a 126 GeV Higgs boson provided the scalar mass fell between approximately $850 - 1400$ GeV and the right-handed neutrino remained very heavy (of order $10^{14}$ GeV). The MPP can instead be imposed at the ``string scale'' of $10^{17}$ GeV in order to facilitate Higgs inflation, which results in somewhat lighter DM at around $400-470$ GeV \cite{Hamada:2014xka}. Ref.~\cite{Kawana:2014zxa} includes three additional right-handed neutrinos (one for each generation) at $10^{13}$ GeV and instead of fixing the MPP condition at $M_{\rm Pl}$ allows the boundary condition energy scale to shift, insisting only that $\lambda=\beta_\lambda=0$ at a single scale. Similarly to the other analyses this finds the DM mass must be of order $770-1050$ GeV. Finally, Ref.~\cite{Kawana:2015aca} investigates a gauged B-L model, and claim that this can accommodate an MPP condition applied at $10^{17}$ GeV, as well as Higgs inflation, by tuning the coupling of the Higgs boson to the new scalar.

We see that applying the Planck scale MPP to the real singlet model requires $\lambda = \lambda_S = k_2 = \beta_{\lambda} = \beta_{\lambda_S} = \beta_{k_2} = 0$. However, this constraint will immediately decouple the new scalar state, and the couplings will not be regenerated by renormalization group running. In other words we revert back to the SM. This seems a serious barrier to the MPP, but is not quite as bad as it appears. Firstly, the MPP itself is somewhat imprecise --- the strong first order phase transition made the particular choice of parameters {\em more likely} but some wriggle-room in these parameters is not unreasonable. (How much wriggle-room is appropriate depends on the UV theory of course.) Furthermore, our calculations themselves are imprecise and include uncertainties. We truncate our $\beta$-functions at two-loops and apply approximations to find the MPP solutions themselves. Therefore, it is more appropriate to ask if the MPP constraints can be approximately applied, i.e.\ $\lambda $, $\lambda_S$, $k_2$, $\beta_{\lambda}$, $\beta_{\lambda_S}$ and $\beta_{k_2}$ should be ``small''.

To investigate if small parameters are compatible with the low energy observations we fix all the quartic scalar couplings at $M_{Pl}$. We perform a scan over Planck scale parameters, allowing $\lambda$, $\lambda_S$ and $|\kappa_2|$ to vary between $0$ and $1$. We also allow $v_S$ or $m_S$ to vary between zero and $2\,$TeV in the broken or Dark Matter phases respectively. We use SARAH 4.12.2 \cite{Staub:2013tta} to calculate the two-loop $\beta$ functions as well as the mass matrices, tadpole equations, vertices and loop corrections we need to calculate mass spectra at low energies; and FlexibleSUSY 2.0.1 \cite{Athron:2017fvs,Athron:2014yba,Allanach:2001kg,Allanach:2013kza} is used to build the spectrum generator needed to get the mass spectrum for each point.

Valid parameter choices must result in a vacuum that is bounded from below up to $M_{Pl}$, so we also require, at all scales, the vacuum stability conditions,
\begin{equation}
	\label{eq:real_vsc}
	\lambda, \lambda_S \geq 0, \qquad \qquad
	\sqrt{\lambda \lambda_S} + k_2 \geq 0.
\end{equation}
We also require dimensionless couplings remain perturbative up to $M_{Pl}$, so,
\begin{equation}
	\label{eq:real_perturbative}
	\lambda, \lambda_S, k_2 \leq \sqrt{4 \pi}.
\end{equation}
We further check vacuum stability using Vevacious \cite{Camargo-Molina:2013qva} which minimises the one-loop effective potential and checks that it is indeed the global minimum. We also require that one of the two scalars of the model is a valid SM Higgs, with mass in the range $124.7$ GeV $\leq m_{h,H} \leq 127.1$ GeV. We allow for a wider range of Higgs masses than the experimental uncertainty as an estimate of the theoretical uncertainty associated with the calculation of the mass spectrum.

These constraints already invalidate much of the parameter space, but we must also apply experimental constraints from the LHC, LEP and Tevatron to ensure they are phenomenologically viable. To this end, we employ HiggsBounds \cite{Bechtle:2013wla} and HiggsSignals \cite{Bechtle:2013xfa}, and further use sHDECAY \cite{Costa:2015llh,Coimbra:2013qq,Butterworth:2010ym} to calculate the total widths and branching ratios for each parameter choice.

In the Dark Matter phase we must also include constraints from the dark matter, using  micrOMEGAS \cite{Belanger:2014vza} to calculate the relic density to compare with the combined WMAP \cite{Hinshaw:2012aka} and Planck \cite{Ade:2015xua} result,
\begin{equation}
	\label{eq:real_DM_relic_density}
	\Omega h^2 = 0.1199 \pm 0.0027.
\end{equation}
A point is excluded if the calculated relic density is greater than $\Omega h^2 + 3 \sigma$ to ensure that a DM candidate does not overclose the universe, but we allow for the possibility that there may be some other contributions to the relic density which we are not taking into account. We also include constraints from dark mattter direct detection that place limits on the spin independent cross section of weakly interacting massive particles (WIMPs) on nucleons. The strongest of those constraints comes from the LUX experiment \cite{Akerib:2016vxi}.

\begin{figure}[tbh]
  \centering
  \includegraphics[width=\textwidth]{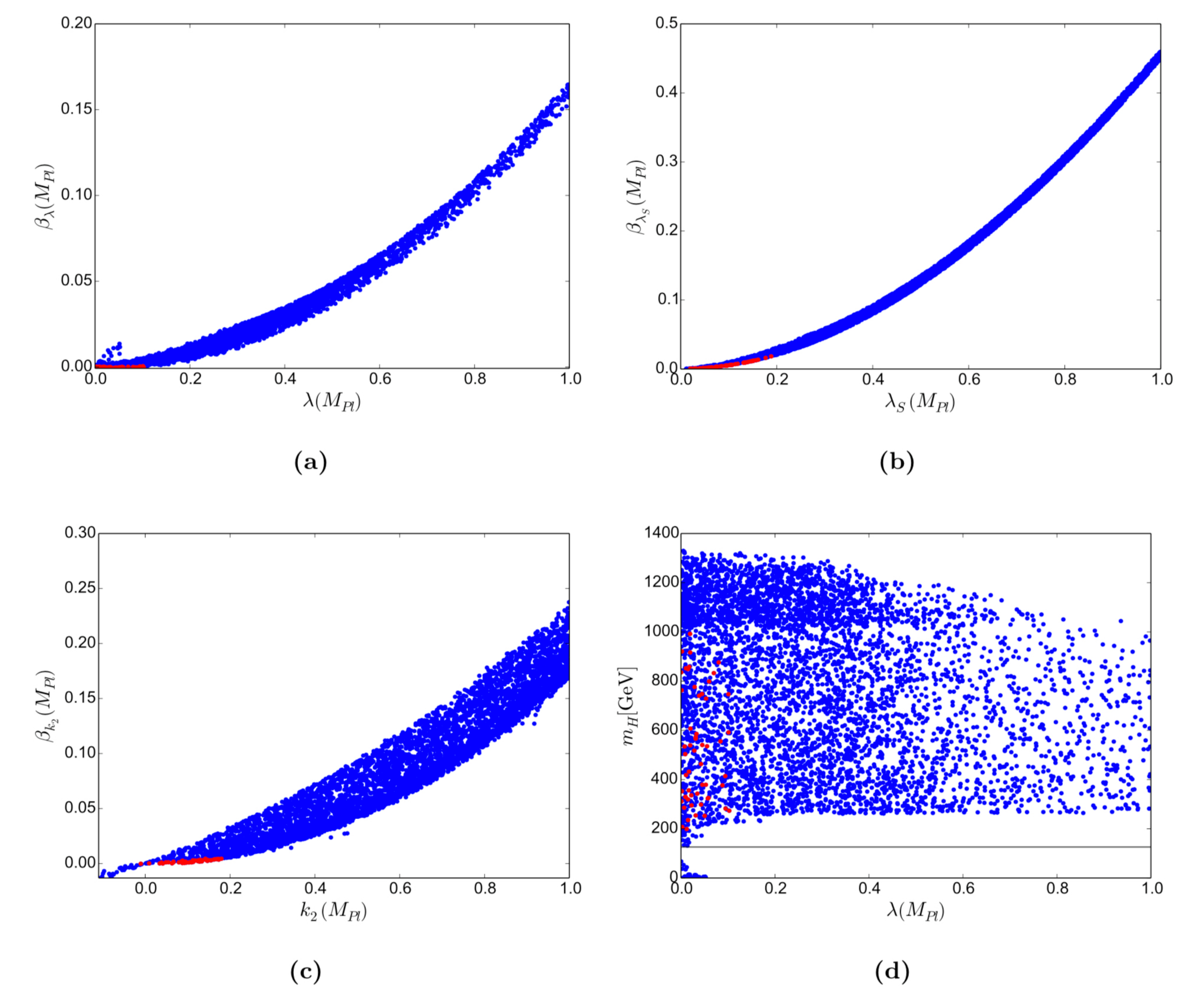}
\caption{Values of \textbf{(a)} $\lambda(M_{\rm Pl})$,
\textbf{(b)} $\lambda_S(M_{\rm Pl})$ and
\textbf{(c)} $\lambda(M_{\rm Pl})$ compared to their respective $\beta$-functions in the broken phase. All points pass theoretical and experimental constraints. Red points further obey $\beta_{\lambda} < 0.0009$, $\beta_{\lambda_S} < 0.019$, $\beta_{k_2} < 0.0045$ at $M_{Pl}$. Also shown \textbf{(d)} is the mass of the additional Higgs for values of $\lambda(M_{\rm Pl})$.}
\label{fig:figure_4}
\end{figure}

We present the results of the analysis of the broken phase in Figure~\ref{fig:figure_4}, where we see lots of parameter choices pass the theoretical and experimental constraints, although only a few of these obey the MPP criterion of the quartic couplings being small. We are interested in points that fall in the lower left corner of Figure~\ref{fig:figure_4}(a-c) as well as those to the left in Figure~\ref{fig:figure_4}(d). To further aid in the discrimination of small values we have coloured red those points for which $\beta_{\lambda} < 0.0009$, $\beta_{\lambda_S} < 0.019$, and $\beta_{k_2} < 0.0045$, which is an estimate of the truncation error in their high scale values as estimated by the difference between the one and two loop Renormalization Group running.

These reasonably numerous red points indicate parameter choices for which there is indeed a second approximately degenerate vacuum at the Planck scale, that provide the correct Higgs boson mass and conform to all low energy observations. This remarkable result need not have been the case. Unfortunately we have also lost predictive power. The SM Higgs mass is fixed by our constraints, so not a prediction and the new Higgs mass can take on a rather wide range of values between $200\,$GeV and $2\,$TeV.

\begin{figure}[tbh]
  \centering
  \includegraphics[width=\textwidth]{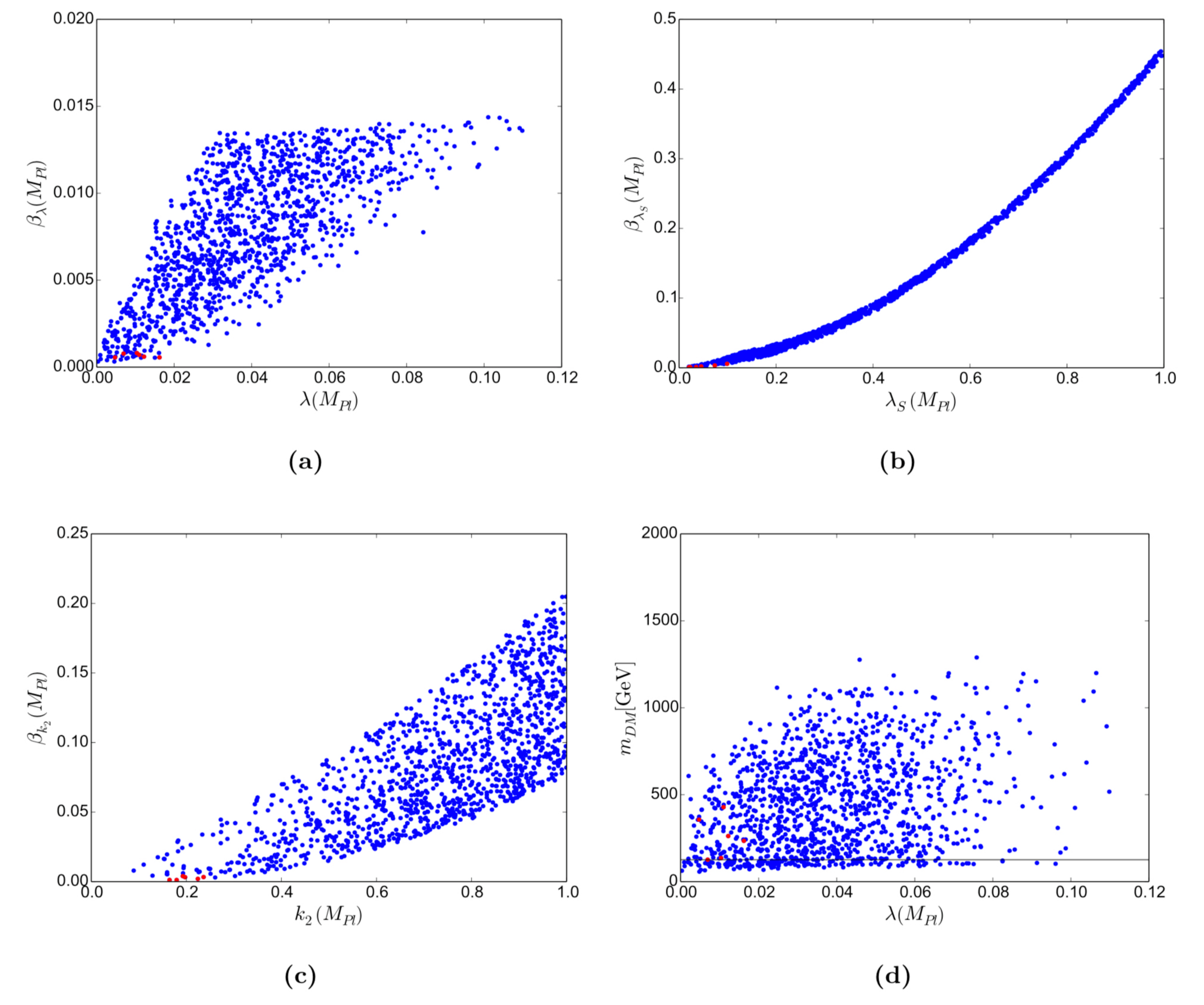}
\caption{Values of \textbf{(a)} $\lambda(M_{\rm Pl})$,
\textbf{(b)} $\lambda_S(M_{\rm Pl})$ and
\textbf{(c)} $\lambda(M_{\rm Pl})$ compared to their respective $\beta$-functions in the Dark Matter phase. All points pass theoretical and experimental constraints. Red points further obey $\beta_{\lambda} < 0.0009$, $\beta_{\lambda_S} < 0.019$, $\beta_{k_2} < 0.0045$ at $M_{Pl}$. Also shown \textbf{(d)} is the mass of the additional scalar for values of $\lambda(M_{\rm Pl})$.}
\label{fig:figure_5}
\end{figure}

It is much more difficult to accommodate the MPP in the Dark Matter phase, as can bee seen in Figure~\ref{fig:figure_5}, which is in part due to the extra constraint from Dark Matter which considerably reduces the acceptable points. We do see parameter choices that evade all constraints with very small values of the $\beta$-functions (red points) but these often have rather large values of the quartic couplings. This is especially true for $\kappa_2$ but is also true, to a lesser degree, for $\lambda$.

\section{A Complex Singlet Extension}
\label{sec:complex_singlet}
We may complicate the model only slightly be promoting our new singlet to a complex field, $\mathbb{S} = S_1 + i S_2$, and consider a potential of the form~\citep{Barger:2008jx, Barger:2010yn, Gonderinger:2012rd, Costa:2014qga, Coimbra:2013qq, Robens:2015gla, Muhlleitner:2017dkd}
\begin{equation}
\label{eq:complex_potential}
V  = \frac{\mu^2}{2} H^{\dagger} H + \frac{\lambda}{4} \left( H^{\dagger} H \right)^2 + \frac{\delta}{2} \left( H^{\dagger} H \right) \left| \mathbb{S} \right|^2 + \frac{b_2}{2} \left| \mathbb{S} \right|^2 + \frac{d_2}{4} \left| \mathbb{S} \right|^4
+ \left( \frac{b_1}{4} \mathbb{S}^2 + a_1 \mathbb{S} + c.c \right).
\end{equation}
For computational convenience we define
\begin{equation}
	b_{\pm} = \frac{1}{2} \left (b_2 \pm b_1 \right),
\end{equation}
which function as the (squared) masses if the model is recast as two real scalar fields. The complex singlet field may acquire a non-zero vev for its real, and possibly imaginary, part. If both real and imaginary parts acquire non-zero vevs,
\begin{equation}
\label{eq:complex_vevs}
\mathbb{S} = \frac{1}{\sqrt{2}} \left[ v_{s_1} + s_1 + i \left( v_{s_2} + s_2 \right) \right],
\end{equation}
we again call this the ``broken phase'' following our earlier nomenclature (introduced in Ref.~\cite{Costa:2014qga}). Therefore, in addition to the bilinear terms $\mu^2$ and $b_{\pm}$ which are fixed via the electroweak vacuum minimisation conditions, the model is described by
\begin{equation}
\label{eq:complex_inputs_broken}
\lambda, \quad d_2, \quad \delta, \quad v_{s_1}, \quad v_{s_2}, \quad a_1.
\end{equation}
In this phase, all three scalar field fluctuations $h$, $s_1$ and $s_2$ mix.

In contrast, if the vev of the imaginary part remains zero, the second electroweak vacuum minimisation condition (for $S_2$) is trivial and $b_-$ becomes a free parameter. In this case the input parameters are
\begin{equation}
\label{eq:complex_inputs_DM}
\lambda, \quad d_2, \quad \delta, \quad v_{s_1}, \quad b_{-}, \quad a_1.
\end{equation}
Now we find ourselves in the ``dark matter phase'', where mixing is allowed  between $h$ and the real part of the complex singlet field $s_1$. The imaginary part $s_2$ does not mix and is a dark matter candidate kept stable by the symmetry $S_2 \to - S_2$.

The numerical analysis of this model follows closely with that of the real singlet extension discussed above.  We scan over $\lambda$, $d_2$ and $\delta$, allowing them to vary between $0$ and $0.5$; $v_{s_1}$ and $v_{s_2}$, if present, are allowed to take values up to $2\,$TeV; $b_-$ has dimension mass$^2$ and is allowed to range to $10^5\,$GeV$^2$. Finally $a_1$, with dimension mass$^3$ and is  allowed up to $10^8\,$GeV$^3$.

We make use of SARAH and FlexibleSUSY again (though slightly older versions, 4.9.3 and 1.6.1 respectively). Constraints on vacuum stability and perturbativity are again applied; in this case stability requires \citep{Gonderinger:2012rd}
\begin{equation}
\label{eq:complex_vacuumstability1}
\lambda, d_2 \geq 0, \qquad \qquad
\delta + \sqrt{\lambda d_2} \geq 0.
\end{equation}
The global minimum is ensured with Vevacious. Finally, we allowed the same Higgs mass range as before and apply experimental constraints using HiggsBounds, and HiggsSignals \cite{Bechtle:2013xfa}, and sHDECAY. MicrOMEGAS is used to provide constraints from Dark Matter in the Dark Matter phase. For further details of this anaylis see Ref.~\cite{McDowall:2018tdg}.

We are in principle interested in the high scale constraints $\lambda = \beta_{\lambda} = 0$, $d_2 = \beta_{d_2} =0$ and $\delta=\beta_\delta=0$. However, similar to the real scalar case, we note that setting $\delta$ to zero at $M_{\rm Pl}$ decouples the extra scalars from the SM, and since $\beta_\delta=0$ for this choice, $\delta$ remains zero at all scales and the new scalars are unobservable. We are therefore forced to only consider $\delta$ ``small''. The situation for $d_2$ is slightly more subtle --- for non-zero values of $\delta$, we cannot set $d_2$ exactly to zero at $M_{\rm Pl}$ since it is immediately driven negative by RG running and the vacuum destabilises according to (\ref{eq:complex_vacuumstability1}). So again, we are forced to only consider $d_2$ ``small'' at the Planck scale and indeed must keep it large enough at $M_{\rm Pl}$ to stop it running negative. Fortunately this is not too onerous, and stability is still viable with $d_2$ as small as $0.005$ at the Planck scale, but it is not really clear how large we should permit this to be and still regard the MPP as ``approximately valid''.

In the broken phase, we now have three neutral scalars that mix. One must provide the SM Higgs, while we will call the other two $m_{h_{\rm Light}}$ and $m_{h_{\rm Heavy}}$. Obviously $m_{h_{\rm Light}} < m_{h_{\rm Heavy}}$, but that $h_{\rm Light}$ may still be heavier than the SM-like Higgs, or correspondingly $h_{\rm Heavy}$ may be lighter. We note that these new states may be considerably lighter than the discovered Higgs mass as long as the component from the doublet is not too large, leaving it relatively decoupled.

\begin{figure}[!tbp]
  \centering
  \includegraphics[width=\textwidth]{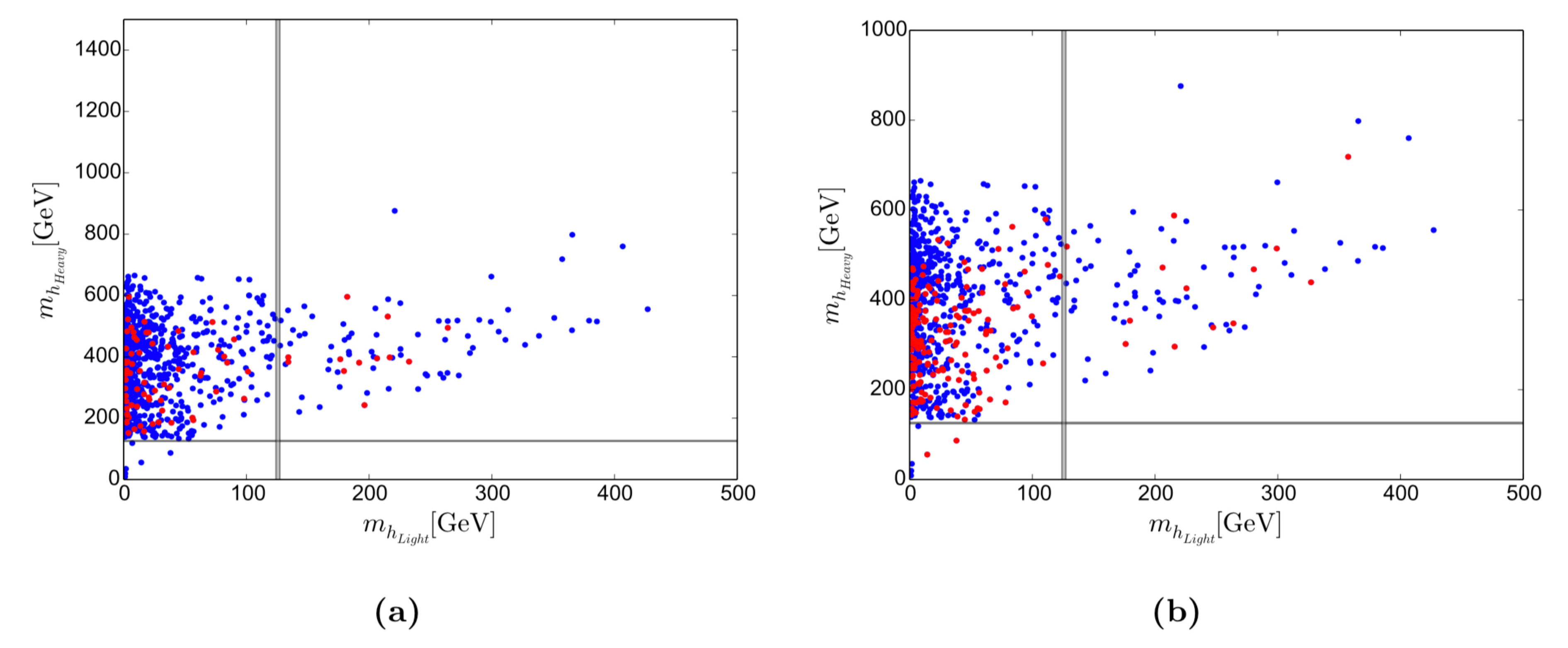}
\caption{Values of $m_{h_{Light}}$ and $m_{h_{Heavy}}$ in the broken phase. All points obey $\lambda<0.067$ and $\beta_{\lambda,\delta,d_2} < 0.05$ at $M_{\rm Pl}$. The grey bands highlight the SM Higgs mass range. \textbf{(a)} Red points obey the more restrictive condition $\beta_\lambda < 0.00005$.
\textbf{(b)} Red points obey $\beta_\lambda < 0.0005$, $\beta_\delta<0.0025$ and $\beta_{d_2}<0.01$. These plots originally appeared in Ref.~\cite{McDowall:2018tdg}.
\label{fig:complex_lighthiggs_heavyhiggs_comparison}}
\end{figure}

This time we will look first at surviving scenarios in the $m_{h_{Light}} - m_{h_{Heavy}}$ plane, with small values of $\lambda$ and $\beta_\lambda$ at the high scale. In Figure~\ref{fig:complex_lighthiggs_heavyhiggs_comparison}(a) we see scenarios that survive all theoretical and experimental constraints. For clarity of the plot, we restrict our points to those with $\lambda<0.067$ and $\beta_{\lambda,\delta,d_2} < 0.05$ at $M_{\rm Pl}$. Points shown in red have been further restricted to have exceptionally small values of $\beta_{\lambda} < 0.00005$, which is the appropriate truncation error arising from the RG running. Corresponding restrictions on $\beta_\delta$ and $\beta_{d_2}$ would be $\beta_\delta<0.00025$ and $\beta_{d_2}<0.001$, but unfortunately we find if we apply these then no points survive.

However, we are reluctant to declare the MPP incompatible with the complex singlet extension. These restrictions on the $\beta$-functions are exceedingly severe and may be too strong. Without knowing the form of the UV completion, we don't know the size of any possible threshold corrections that might arise was we approach the Planck scale, so really don't know how much deviation from zero we should allow in our boundary conditions. To allow some extra slack, we can somewhat arbitrarily relax our boundary condition $\beta$-function cut-offs to ten times the truncation error. We now find some points survive and plot these in figure \ref{fig:complex_lighthiggs_heavyhiggs_comparison}(b). Notice that a small number of points survive that have the SM Higgs as the heaviest of the three scalars.

\begin{figure}[!tbp]
  \centering
	\includegraphics[width=\textwidth]{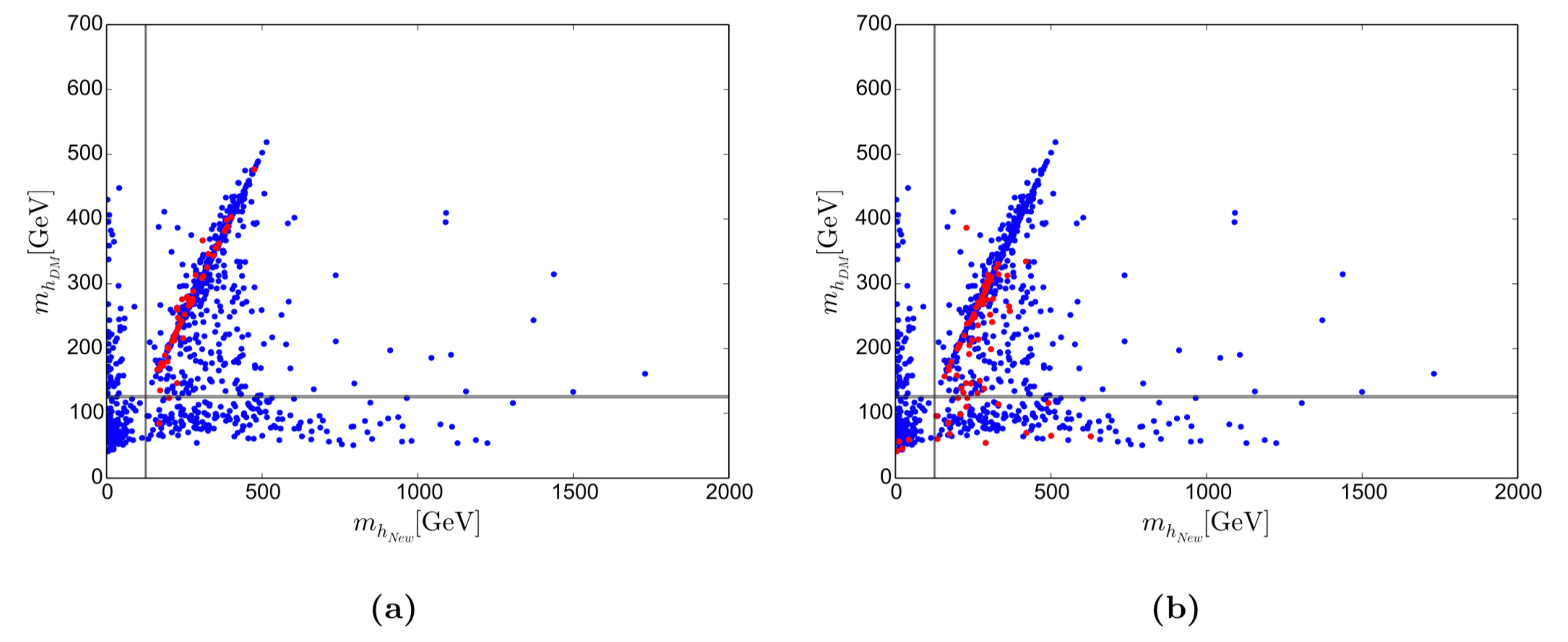}
\caption{Values of $m_{h_{Light}}$ and $m_{h_{Heavy}}$ in the DM phase. All points obey $\lambda<0.067$ and $\beta_{\lambda,\delta,d_2} < 0.05$ at $M_{\rm Pl}$. The grey bands highlight the SM Higgs mass range. \textbf{(a)} Red points obey the more restrictive condition $\beta_\lambda < 0.00005$.
\textbf{(b)} Red points obey $\beta_\lambda < 0.0005$, $\beta_\delta<0.0025$ and $\beta_{d_2}<0.01$. These plots originally appeared in Ref.~\cite{McDowall:2018tdg}.
\label{fig:complex_DM_additionalhiggs_DMhiggs_comparison}}
\end{figure}

In the dark matter phase only two of the three scalars are allowed to mix, with the third becoming a dark matter candidate. We call the non-SM-like Higgs $h_{\rm New}$ whilst the DM scalar is $h_{DM}$. Figure \ref{fig:complex_DM_additionalhiggs_DMhiggs_comparison}(a) examines these extra scalar masses when we restrict $\lambda$ and $\beta_\lambda$ to be consistent with zero.  Again, for clarity of the plot, we show only points with  $\beta_\lambda<0.05$ in blue before demonstrating the effect of the constraint $\beta_\lambda<0.00005$ in red. It is interesting to note that no points with $m_{h_{New}} < m_{h_{SM}}$ survive the stronger constraint on $\beta_{\lambda}$, and the majority of the points that do survive have almost degenerate masses of $m_{h_{New}}$ and $m_{h_{DM}}$. The tree level masses of $m_{h_{New}}$ ($m_{h_{DM}}$) have a linear dependence on $a_1$ ($b_{-}$) which appears to dominate when both of the additional scalars are heavier than the SM Higgs.

Figure \ref{fig:complex_DM_additionalhiggs_DMhiggs_comparison}(a) might suggest that small values of the $\beta$ functions at the Planck scale correlates with a small mass difference $\Delta m = |m_{h_{New}} - m_{h_{DM}}|$. However, while $80\%$ of the points that pass through the constraint $\lambda<0.067, \beta_\lambda < 0.00005$ result in $\Delta m < 40$ GeV, so do $67\%$ of the points that don't. This tendency towards degeneracy is a feature of all of the points that satisfy the theoretical constraints. These points exhibit small values of the soft $U(1)$ breaking parameters $a_1$ and $b_1$, forcing a small $\Delta m$ \citep{Coimbra:2013qq}. It is interesting to note that many points in the degenerate mass region can completely account for the dark matter relic density. The degeneracy opens up co-annihilation channels involving both $m_{h_{DM}}$ and $m_{h_{New}}$ that enter the relic density calculation \citep{Baker:2015qna,Ghorbani:2014gka}. These new channels help bring down the relic density to within the $3 \sigma$ range.

As in the broken phase, no DM phase points survive when the severe truncation error cut-offs are applied simultaneously with the experimental constraints.
However, we see scenarios survive if we relax the constraints by a factor of 10. These scenarios are shown in Figure \ref{fig:complex_DM_additionalhiggs_DMhiggs_comparison}(b).

\section{The Two Higgs Doublet Model}
\label{sec:THDM}

Finally we will examine models with two Higgs doublets to see if they are compatible with the MPP. The most general potential of the Two Higgs Doublet Model (2HDM) (see Ref.~\cite{Branco:2011iw} for a useful review) is,
\begin{eqnarray}
	V \left(H_1, H_2 \right) &=& m_{11}^2 H_{1}^{\dagger} H_{1} + m_{22}^2 H_{2}^{\dagger} H_{2} - \left( m_{12}^2 H_{1}^{\dagger} H_{2} + c.c \right)  + \lambda_1 \left( H_{1}^{\dagger} H_{1} \right)^2 \\ \nonumber
	&&+ \lambda_2 \left( H_{2}^{\dagger} H_{2} \right)^2 + \lambda_3 \left( H_{1}^{\dagger} H_{1} \right) \left( H_{2}^{\dagger} H_{2} \right) + \lambda_4 \left( H_{1}^{\dagger} H_{2} \right) \left( H_{2}^{\dagger} H_{1} \right) \\ \nonumber
	&&+ \left( \frac{\lambda_5}{2} \left( H_{1}^{\dagger} H_{2} \right)^2 + \lambda_6 \left( H_{1}^{\dagger} H_{1} \right) \left( H_{1}^{\dagger} H_{2} \right) + \lambda_7 \left( H_{2}^{\dagger} H_{2} \right) \left( H_{1}^{\dagger} H_{2} \right) + c.c \right),
	\label{eq:THDM_potential}
\end{eqnarray}
where the two Higgs-doublets themselves are given by,
\begin{equation}
	H_n = \begin{pmatrix} \chi_n^{+} \\ \left( H_n^0 + i A_n^0 \right)/\sqrt{2} \end{pmatrix}, \quad n=1,2.
	\label{eq:THDM_Hn}
\end{equation}

The parameters $m_{11}^2$, $m_{22}^2$ and $\lambda_{1,2,3,4}$ are real, whilst $m_{12}^2$ and $\lambda_{5,6,7}$ can in princple be complex and induce CP violation. During electroweak symmetry breaking the neutral components of the Higgs fields, $H_n^0$, develop vacuum expectation values (vevs) $\langle H_n^0 \rangle = v_n / \sqrt{2}$. The relationship to the SM vev $v = \sqrt{v_1^2 + v_2^2} = 246\, \mathrm{GeV}$ is determined by the Fermi constant but the ratio of the vevs, $\tan \beta = v_2/v_1$, is a free parameter. The physical scalar sector of the model includes two neutral scalar Higgs $h$ and $H$, a pseudoscalar Higgs $A$ and the charged Higgs $H^{\pm}$.

It's clear that the 2HDM potential is considerably more complicated than its Standard Model counterpart, so it's common to employ additional global symmetries to increase the predictivity of the model. There are only six possible types of global symmetry that have a distinctive effect on the potential \citep{Ivanov:2006yq,Ferreira:2015rha}. The 2HDM has been considered for suitability of the MPP in Refs.~\cite{Froggatt:2004st, Laperashvili:2004bu,Froggatt:2006zc,Froggatt:2007qp,Froggatt:2007py,McDowall:2018ulq}, though all but the last of these predate the Higgs discovery so could not be confronted with the measured Higgs mass. Ref.~\cite{Froggatt:2007py} is notable in that it shows that the MPP itself may be used as a mechanism for suppressing CP-violabtion and Flavour Changing Neutral Currents (FCNCs).

In Ref.\cite{McDowall:2018ulq} we took the more usual route of implementing a $\mathbb{Z}_2$ symmetry to forbid FCNCs by allowing only one type of fermion to couple to one Higgs doublet. This requirement sets $\lambda_6$, $\lambda_7$ and $m_{12}$ to zero. Following Ref.\cite{McDowall:2018ulq}'s treatment, we may then softly break this $\mathbb{Z}_2$ by re-introducing a (real) non-zero $m_{12}$. We will restrict ourselves to a Type-II model where up-type quarks and leptons couple to the first Higgs-doublet and down-type quarks to the second Higgs-doublet, though we note that the most significant effect of the Yukawa sector comes from which doublet the top-quark couples to, so results for other 2HDM Yukawa assignments would be very similar to those for Type-II.

For each parameter point the model is described by the bilinear terms $m_{11}$ and $m_{22}$, which are replaced by $M_Z$ and $\tan \beta$ by applying the electroweak vacuum minimisation conditions, as well as the additonal input parameters, $m_{12}$ and $\lambda_{i}(M_{\rm Pl})$ with $i=1\ldots 5$. As previously we use SARAH to calculate the two-loop $\beta$ functions, which are used by FlexibleSUSY to run the couplings between $M_Z$ and $M_{\rm Pl}$.

We also consider a simpler model, the Inert Doublet Model (IDM), where we introduce an additional unbroken $\mathbb{Z}_2$ symmetry, under which the new doublet has odd parity but all other fields are even (see \cite{Ilnicka:2015jba} for a useful review). The scalar sector now consists of the SM Higgs field $H$ and an inert doublet $\Phi$, with mixing between the two forbidden by the new symmetry. The inert doublet does not couple to any of the SM fields and does not gain a vacuum expectation value.

The potential is,
\begin{eqnarray}
	V \left(H, \Phi \right) &=& m_{11}^2 H^{\dagger} H + m_{22}^2 \Phi^{\dagger} \Phi + \lambda_1 \left( H^{\dagger} H \right)^2 + \lambda_2 \left( \Phi^{\dagger} \Phi \right)^2 \\ \nonumber
	&&+ \lambda_3 \left( H^{\dagger} H \right) \left( \Phi^{\dagger} \Phi \right) + \lambda_4 \left( H^{\dagger} \Phi \right) \left( \Phi^{\dagger} H \right) + \left( \frac{\lambda_5}{2} \left( H^{\dagger} \Phi \right)^2 + c.c \right).
	\label{eq:Inert_potential}
\end{eqnarray}
Once again the quartic coupling can have complex values, but we will focus on the real-valued case. Note that now the mixing term proportional to $m_{12}^2$ is absent. During electroweak symmetry breaking the neutral component of the SM Higgs doublet acquires a vacuum expectation value $v = 246\,$GeV. The neutral Higgs $h$ corresponds to the SM Higgs boson whilst $H$, $A$ and $H^{\pm}$ are inert scalars. The lightest of these is stable thanks to the $\mathbb{Z}_2$ symmetry and, assuming it is one of the neutral scalars $H$ or $A$, it is a potential Dark Matter (DM) candidate \citep{PhysRevD.95.015017,PhysRevD.92.055006}.

As in the previous case, the mass term associated with the SM Higgs doublet $m_{11}^2$ is fixed via the electroweak minimisation conditions, but now we don't have a second vev to fix $m_{22}^2$, which must remain an input. Our input parameters are therefore $m_{22}$ and $\lambda_{i}(M_{\rm Pl})$ with $i=1\ldots 5$. As in the Type-II model, we use SARAH and FlexibleSUSY to calculate the mass spectrum and to run couplings between the low and high scales of interest.

Valid points in our parameter space scan must be perturbative up to the Planck scale. For the Higgs quartic couplings this requires them to satisfy $\lambda_i < \sqrt{4 \pi}$ up to $M_{Pl}$. We require points that are bounded from below at all scales up to $M_{Pl}$ \citep{Chowdhury:2015yja}. To that end we check if the boundedness conditions \citep{Branco:2011iw},
\begin{eqnarray}
\lambda_1,
\lambda_2 &>& 0, \\ \nonumber
\lambda_3 &>& -2 \sqrt{\lambda_1 \lambda_2}, \\ \nonumber
\lambda_3 + \lambda_4 - |\lambda_5| &>& -2 \sqrt{\lambda_1 \lambda_2},
\label{eq:THDM_VSCs}
\end{eqnarray}
are met at all scales \citep{Sher:1988mj,Chataignier:2018aud}.

\begin{figure}[htb!]
  \centering
	\includegraphics[width=\textwidth]{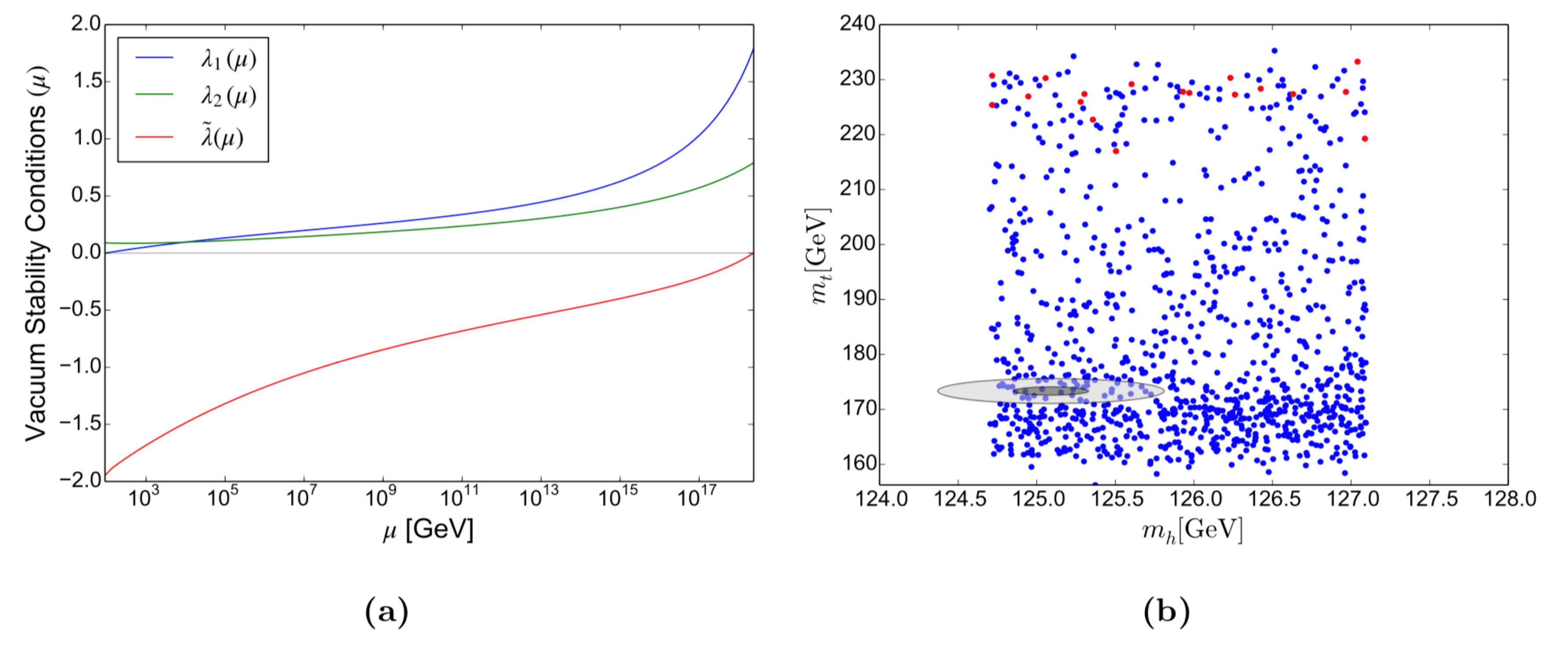}
\caption{\textbf{(a)} Example running of $\lambda_1$, $\lambda_2$ and $\tilde{\lambda}$ for a point that provides valid masses for the SM Higgs and the top quark in the Type-II Two Higgs Doublet Model. Boundedness from below and vacuum stability requires that all three couplings are positive at all scales. \textbf{(b)} Results of our Multiple Point Principle scan in the $m_{h} - m_{t}$ plane of the Type-II Two Higgs Doublet Model. The blue points provide valid SM higgs masses whilst the red points also pass the vacuum stability conditions at all scales. The ellipses show the experimentally allowed values of $m_t$ and $m_{h}$ at $1 \,\sigma$ (dark grey) and $3 \,\sigma$ (light grey) uncertainty. These plots originally appeared in Ref.~\cite{McDowall:2018ulq}.
\label{fig:THDMII_MPP_comparison}}
\end{figure}

The goal for the MPP is to have an additional minimum at $M_{\rm Pl}$, degenerate with the electroweak minimum,. This is naively satisfied if all of the quartic couplings are zero at $M_{\rm Pl}$, i.e.\ $\lambda_i = 0, i = 1 \ldots 5$. However, the RG running of $\lambda_1$ and $\lambda_2$ results in an unstable vacuum configuration \citep{Froggatt:2004st, Laperashvili:2004bu,Froggatt:2006zc,Froggatt:2007qp}. It is also possible for degenerate vacua to exist within the 2HDM if we relax the condition $\lambda_i = 0$. Specifically, by allowing $\lambda_1$, $\lambda_2$, $\lambda_3$ and $\lambda_4$ to be non-zero at $M_{\rm Pl}$, the following conditions \citep{Froggatt:2004st} are consistent with the implementation of the MPP at $ M_{\rm Pl}$;
\begin{eqnarray}
\label{eq:THDM_MPP_conditions}
\lambda_5 \left( M_{\rm Pl} \right) &=& 0 \\ \nonumber
\lambda_4 \left(  M_{\rm Pl} \right) &<& 0 \\ \nonumber
\tilde{\lambda} \left(  M_{\rm Pl} \right) = \sqrt{\lambda_1 \lambda_2} + \lambda_3 + \text{min}(0,\lambda_4) &=& 0 \\ \nonumber
\beta_{\tilde{\lambda}} \left(  M_{\rm Pl} \right) &=& 0.
\end{eqnarray}
To investigate whether these MPP conditions in the Type-II 2HDM are consistent with the current experimental constraints on the SM Higgs mass $m_h$ and the top-quark mass $m_t$, we generated points in the parameter space, applying the theoretical constraint of vacuum stability at all scales. Figure \ref{fig:THDMII_MPP_comparison}(a) shows an example of the running of $\lambda_1$, $\lambda_2$ and $\tilde{\lambda}$ for a point that results in experimentally valid values of the SM Higgs mass and the top-quark mass, and is also consistent with the MPP conditions of (\ref{eq:THDM_MPP_conditions}). Vacuum stability requires that all of these couplings remain greater than zero at all scales, but the negative running of $\tilde{\lambda}$ pulls it to negative values.

Figure \ref{fig:THDMII_MPP_comparison}(b) shows an investigation of the $m_h - m_t$ plane, where we temproarily suspend vacuum stability to demonstrate the effect. We see plenty of valid points in blue, where vacuum stability is not required. However, the points that satisfy the vacuum stability conditions, highlighted in red, have larger values of the top Yukawa $y_t$ which positively contribute to the running of the quartic couplings. The larger required $y_t$ corresponds to a top mass in the range $220 \lesssim m_t \lesssim 230\,$GeV which is not compatible with current experimental bounds on the top-quark mass.

These MPP constraints also apply to the Inert Doublet Model. We examined the IDM parameter space as we did for the Type-II 2HDM, applying the MPP conditions at $M_{Pl}$ and requiring valid points to be stable up to the Planck scale and to have a SM Higgs candidate.

\begin{figure}[htb!]
  \centering
\includegraphics[width=0.5\textwidth]{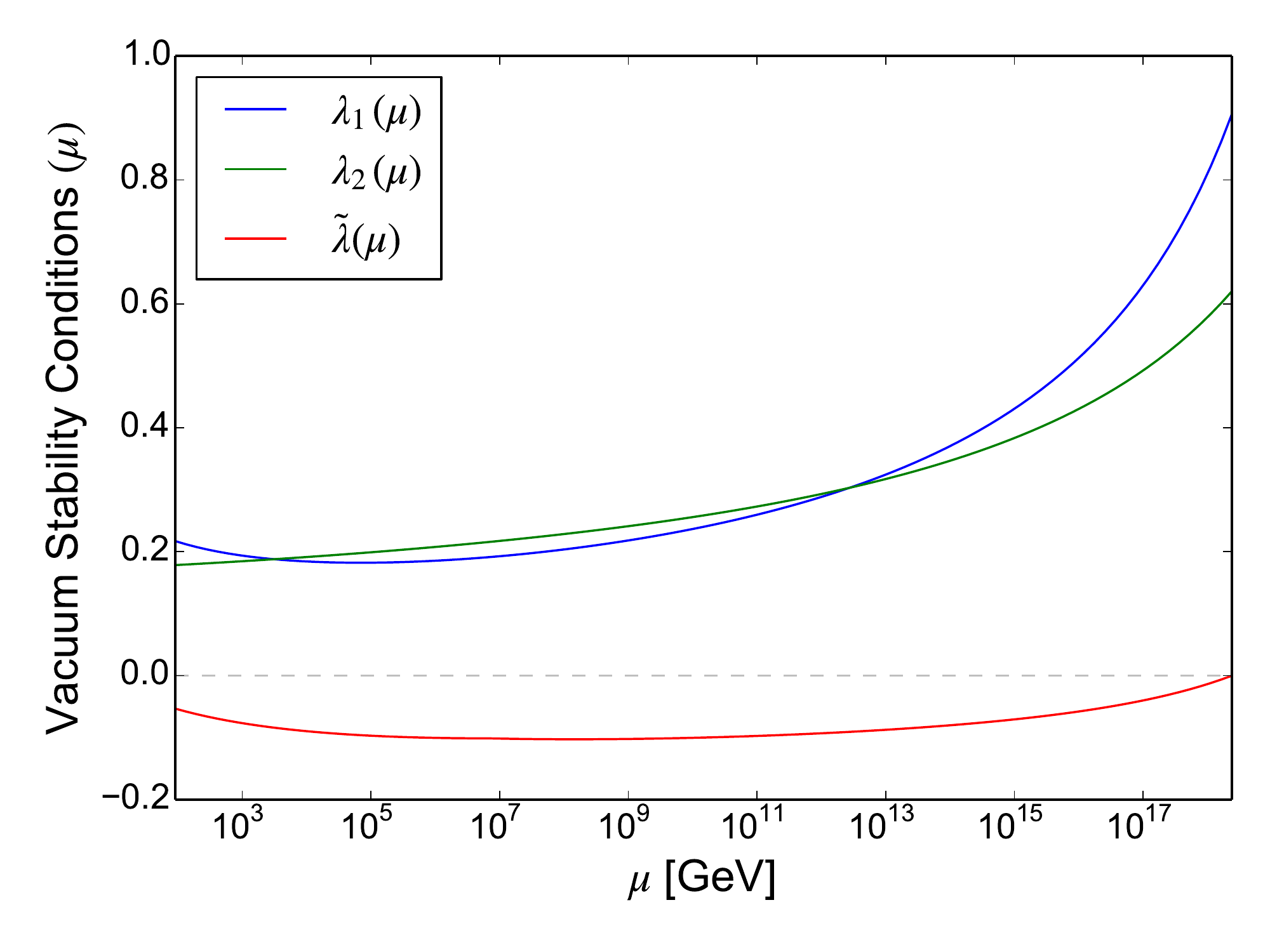}
\caption{Example running of $\lambda_1$, $\lambda_2$ and $\tilde{\lambda}$ for a point that provides valid masses for the SM Higgs and the top quark in the Inert Doublet Model. Boundedness from below and vacuum stability requires that all three couplings are positive at all scales. This plot originally appeared in Ref.~\cite{McDowall:2018ulq}.
\label{fig:Inert_MPP_vsc}}
\end{figure}

Figure \ref{fig:Inert_MPP_vsc} shows the running of the quartic couplings $\lambda_1$, $\lambda_2$ and  $\tilde{\lambda}$ for an example point in our scan that provided a valid SM Higgs and top mass. As in the Type-II model, a stable vacuum requires all three of these couplings to be positive at all scales. Clearly this point fails our vacuum stability test, and unfortunately it is representative of the other points in our scan. We found {\em no points} that could simultaneously satisfy the constraints of perturbativity, vacuum stability and the requirement of a realistic SM mass spectrum. Specifically, there are points that provide valid SM Higgs and top masses, but all of these points fail the condition $\tilde{\lambda} > 0$. In fact, we found no points that could satisfy the MPP conditions outlined in (\ref{eq:THDM_MPP_conditions}) that remained stable up to the Planck scale, regardless of their Higgs or top masses. This therefore suggests that the multiple point principle cannot be implemented successfully in the Inert Doublet Model.

\section{More Exotic Models}

The MPP has also been applied to several other models of new physics, of varying degrees of complexity. For example, Ref.~\cite{Hamada:2015fma} consider one of the more minimal extensions by including either a Majorana fermion triplet or a real scalar triplet, and in both cases were able to find good agreement with the MPP by keeping the new states rather heavy (of order $10^{16}$ GeV for the fermion triplet and slightly higher for the scalar).

Ref.~\cite{bennett:1997kq} studies what the authors term an ``anti-GUT'' within the context of the SM. This is a model where each generation comes with a full complement of the SM gauge groups, augmented with an additional local $U(1)$, so that the full group (at high energies) is $[SU(3) \times U(2) \times U(1)]^3 \times U(1)$. The resulting Higgs mass prediction is $139 \pm 16$ GeV, though the uncertainty in this prediction would no doubt be significantly reduced with more modern inputs, and they also find reasonable agreement with the SM Yukawa couplings.

Another proposed alternative is to mix a fundamental scalar with the scalar bound states of a new strongly interacting gauge symmetry~\cite{haba:2017quk}. This allows for the dynamical generation of the Higgs mass, with a classically scale invariant theory satisfying the MPP condition. They predict new scalar states at approximately 300 GeV as well as a new gauge boson coupling to the SM fermions.

The MPP may also be used to constrain theories with extra dimensions. Ref.~\cite{Hamada:2017yji} examines the SM compactified at high scales onto $S^1$ and $T^1$, additionally applying the MPP. They find this constrains the neutrinos in the model to have Dirac masses, with the lightest of order $1-10$ meV. This would prevent neutrinoless double-beta decay and have interesting cosmological consequences.

An more exotic suggestion comes from the original authors of the MPP: the existence of a bound state made of six top-quarks and six anti-top-quarks~\citep{Froggatt:2004nn,Froggatt:2004bh,Das:2008an,Laperashvili:2016cah,Laperashvili:2017sih,Nielsen:2017ows}. They postulate a new phase different from and degenerate with the standard electroweak Higgs phase, caused by the condensation of this new top-anti-top bound state. They claim this bound state arises from the exchange of Higgs bosons due to the large top Yukawa coupling. Therefore the MPP is extended to insist on not just two, but three degenrate vacua: two at low energies and one at the Planck scale. The authors also claim that the extra energy density of this new bound state provides a solution for the cosmological constant problem.

\section{Summary and Conclusions}
\label{sec:conclusions}

The measured value of the Higgs boson mass implies that, if the SM is true to high scales, the Higgs quartic coupling and its $\beta$-function are intriguingly close to zero at the Planck scale. Indeed, their values imply that the SM vacuum is metastable, with a slightly deeper vacuum at the Planck scale.

One suggested explanation for this is the {\em Multiple Point Principle}. By considering {\em extensive} variables, nature tends to choose Higgs parameters so that different phases of Electroweak symmetry breaking may coexist. This predicts a second degenerate vacuum at the Planck scale, rather similar to that implied by the Higgs measurements. An analysis of the MPP in the SM provides a prediction of the Higgs mass, $m_h = 129 \pm 1.5\,$Gev which is slightly above the measured value. It is therefore intersting to ask how extensions to the SM might change this picture, especially since we do expect new physics to appear well before the Planck scale. In this paper, we have reviewed the compatibility of the MPP with simple Higgs sector extensions, considering both extra scalars and doublets.

We began our review of extended models by considering an additional real scalar field, in both the broken and Dark Matter phases. We had to weaken the MPP constraints somewhat in order to prevent the extra states from decoupling, but found promising results. These real scalar extensions were both compatible with the (relaxed) MPP, though working scenarios in the Dark Matter case were rare due to the additional Dark Matter constraints. Unfortunately the MPP didn't prove very predictive because it left us with a wide range of allowed additional scalar masses.

The next extension we considered was an extra complex singlet, where again we had to relax the MPP condition in order to prevent decoupling. We also found that we were unable to keep the parameter $d_2$ very small at the Planck scale since it tended to run negative, destabilizing the vacuum. Furthermore, our constraints setting the $\beta$-functions for the Higgs parameters to zero could not all be accommodated simultaneously while keeping viable low energy phenomenology. However, relaxing these constraints somewhat did again yield scenarios that are stable, evade experimental constraints, have the correct Higgs mass, and in the Dark Matter phase, provide the correct relic density,

Finally we investigated the Type-II Two Higgs Doublet Model and the Inert Doublet Model. Models with a second Higgs doublet have much more flexibility in their scalar potential, which one might expect gives them more freedom to accommodate the boundary conditions of the the MPP. However, we found that both the Type-II 2HDM and the IDM cannot satisfy the conditions required at the Planck scale by the MPP. Specifically, we found no points in either model's parameter space that was consistent with the MPP whilst also having a valid SM Higgs, an experimentally acceptable top quark mass, and a stable vacuum. In the Type-II case we found that a stable vacuum would require a top mass on the order of $230\,$GeV, whilst in the Inert case we found no points at all that could meet our theoretical requirements. The results of our analysis would suggest that the Multiple Point Principle is not compatible with the Two Higgs Doublet Models that we investigated.

In general it seems rather difficult to accommodate an {\em exact} MPP in any of these models. There are several possible explanations for this. Firstly the MPP conditions may only hold approximately. The original conjecture that there should be a second degenerate vacuum at the Planck scale was itself based on general arguments, and may be realised with some slight modifications. Indeed, one might expect threshold corrections for the new theory to become significant as we approach the Planck scale, slightly modifying the RG running. Secondly, we do expect new physics before the Planck scale to solve the many deficiencies of the SM. It could be that this new physics alters the Higgs running sufficiently to allow the MPP to hold more exactly.  It would be interesting to examine the SM Higgs sector with alternative additions, such as vector-like fermions. Finally, the literature thus far has entirely neglected finite temperature effects in the study of the MPP. Such effects could very well alter the vacuum structure,

Utimately the question remains, is the peculiar behaviour of the SM Higgs potential at the Planck scale a coindindence or a sign of new physics?

\section*{Funding}
DJM acknowledges partial support from the STFC grants ST/L000446/1 and ST/P000746/1.

\section*{Acknowledgments}
The authors would like to thank Peter Athron for invaluable help with FlexibleSUSY; as well as Karl Nordstrom, Ant\'onio Morais and David Sutherland for useful discussions.

\bibliographystyle{frontiersFPHY}
\bibliography{bibliography}

\end{document}